\documentclass[sigconf,10pt]{acmart}
\usepackage{graphicx} 
\usepackage{subcaption}
\usepackage[utf8]{inputenc}
\usepackage[T1]{fontenc}
\usepackage[autostyle=true, english=american]{csquotes}
\usepackage[english]{babel}
\usepackage[capitalize, noabbrev, nameinlink]{cleveref}
\usepackage{enumitem}
\usepackage{microtype}

\date{\today}

\begin{document}

\author{Mathis Engelbart}
\affiliation{%
    \institution{Technical University of Munich}
    \country{Germany}
}
\email{mathis.engelbart@tum.de}

\author{Mike Kosek}
\affiliation{%
    \institution{Technical University of Munich}
    \country{Germany}
}
\email{kosek@in.tum.de}

\author{Lars Eggert}
\affiliation{%
    \institution{Mozilla}
    \country{Finland}
}
\email{leggert@mozilla.com}

\author{Jörg Ott}
\affiliation{%
    \institution{Technical University of Munich}
    \country{Germany}
}
\email{ott@in.tum.de}

\title[From req/res to pub/sub: Exploring Media over QUIC Transport for DNS]{From req/res to pub/sub: Exploring\\Media over QUIC Transport for DNS}

\begin{abstract}
The DNS is a key component of the Internet.
Originally designed to facilitate the resolution of host names to IP addresses, its scope has continuously expanded over the years, today covering use cases such as load balancing or service discovery.
While DNS was initially conceived as a rather static directory service in which resource records (RR) only change rarely, we have seen a number of use cases over the years where a DNS flavor that isn't purely based upon requesting and caching RRs, but rather on an active distribution of updates for all resolvers that showed interest in the respective records in the past, would be preferable.
In this paper, we thus explore a publish-subscribe variant of DNS based on the Media-over-QUIC architecture, where we devise a strawman system and protocol proposal to enable pushing RR updates.
We provide a prototype implementation, finding that DNS can benefit from a publish-subscribe variant: next to limiting update traffic, it can considerably reduce the time it takes for a resolver to receive the latest version of a record, thereby supporting use cases such as load balancing in content distribution networks. The publish-subscribe architecture also brings new challenges to the DNS, including a higher overhead for endpoints due to additional state management, and increased query latencies on first lookup, due to session establishment latencies.
\end{abstract}

\ccsdesc[300]{Networks~Network protocol design}

\acmYear{2025}\copyrightyear{2025}
\setcopyright{cc}
\setcctype[4.0]{by}
\acmConference[HotNets '25]{The 24th ACM Workshop on Hot Topics in Networks}{November 17--18, 2025}{College Park, MD, USA}
\acmBooktitle{The 24th ACM Workshop on Hot Topics in Networks (HotNets '25), November 17--18, 2025, College Park, MD, USA}
\acmDOI{10.1145/3772356.3772416}
\acmISBN{979-8-4007-2280-6/25/11}

\keywords{Media over QUIC, DNS, Publish-Subscribe}

\maketitle

\section{Introduction}
\label{sec:intro}

Since its introduction in the mid-1980s, the Domain Name System (DNS) has been one of the key components of the Internet, replacing the \texttt{hosts.txt} file. Originally designed to support the resolution of host names to IP addresses, aliasing (CNAME), and determining mail and name servers, its scope has continuously expanded to cover protocol and service lookup and load balancing~\cite{rfc2782}, flexible lookup delegation~\cite{rfc3401}, phone number mapping~\cite{rfc6116}, and certificate pinning~\cite{rfc8659}, among other uses.

As such, DNS is effectively a large, global, hierarchically distributed database, consisting of multiple layers where each layer performs specific functions to resolve a name.
\textit{Stub resolvers} are part of every operating system, providing a local interface for DNS requests.
Stub resolvers forward requests to \textit{recursive resolvers} which look up DNS records recursively: starting with requesting the Top-Level Domain (TLD) server for the requested domain from one of the \textit{root nameservers}, then requesting the \textit{authoritative nameserver} from the returned record, and finally requesting the desired record from the identified authoritative nameserver.
Recursive resolvers are typically provided by the ISP (Internet Service Provider) that connect a user or enterprise to the Internet.

DNS was originally designed to be carried over UDP and TCP, of which only UDP was initially globally supported. Secure DNS transports emerged to mitigate eavesdropping and response manipulation~\cite{rfc7626} between stub and recursive resolver, leading to DNS-over-TLS (DoT)~\cite{rfc7858}, DNS-over-HTTPS (DoH)~\cite{rfc8484}, and most recently, DNS-over-QUIC (DoQ)~\cite{rfc9250}. Most prominently, these secure DNS transports are used by public recursive resolvers, such as \texttt{1.1.1.1} (Cloudflare), \texttt{8.8.8.8} (Google), or \texttt{9.9.9.9} (Quad9). They intend to improve user privacy, filter attempts to access malicious sites, or help in accessing otherwise censored content, compared to only using the traditional recursive resolvers provided by ISPs for their customers.
As such, communication between recursive resolvers and authoritative nameservers is still almost exclusively performed unencrypted using DNS over UDP, with DNS over TCP as a fallback for larger responses.

Every layer within the DNS hierarchy operates its own cache where the lifetime of each record is determined by the Time to live (TTL) attributed by the authoritative server.
While typical default configurations use a TTL of 300 seconds, TTLs can be as low, or as high, as deemed appropriate by the provider of the authoritative server; as such, typical TTLs as observed in the Internet range from 10 seconds to 1 day.
While this caching effectively reduces latency and minimizes excess traffic, in particular on aggregation points like recursive resolvers which are used by a multitude of stub resolvers, it also impacts how up-to-date records are: a record is requested from the next layer within the hierarchy only on cache misses, i.e., when the TTL has expired. Thus, in the worst case, a record is as old as the number of caches required for the lookup multiplied by the TTL of the record.

DNS was initially conceived as a rather static directory service in which resource records (RRs) only change rarely and scaled well over its more than 40 years of existence. Yet, DNS has seen a number of uses in which a more dynamic system would be preferable:
For example, Dynamic DNS has been used by (home) users with dynamically assigned IP addresses to run servers in their (home) networks. To cope with changing RRs, an indirection via a DNS anchor service is introduced that supports potentially frequent updates whenever the assigned IP address changes. 
Moreover, Content Distribution Networks (CDNs) such as Akamai use DNS in conjunction with short cache lifetimes for load balancing to redistribute user requests to different servers as a function of their current load\cite{10.1145/1159913.1159962}.
Recently, the IETF TIPTOP (Taking IP to Other Planets) WG began exploring how to extend the Internet architecture into deep space~\cite{many-tiptop-usecase-03,many-tiptop-ip-architecture-01}\footnote{As a possible alternative to the DTN architecture~\cite{rfc4838} and protocol~\cite{rfc9171}.}. One of the most critical issues is coping with long propagation delays at inter-planetary distances since many Internet protocols rely on handshakes.  This also applies to the up front name resolution using DNS, for which active replication of RRs of the relevant domains is proposed as one option~\cite{many-tiptop-dns-00}.

All three scenarios would benefit from a DNS flavor that isn't purely based upon requesting and caching RRs but rather on an active distribution of updates all the way to the last-hop and even stub resolvers---for those resolvers that showed interest in the respective records in the past. This would support spreading updated records for all of the above scenarios at the individual time scales as needed for each of the domains and resources while limiting update traffic.

In this paper, we explore a publish-subscribe variant of DNS that supports incremental deployment.  We begin with a look at the dynamics of today's DNS in §\ref{sec:ttls}. We then leverage a recent development for scalable content distribution that offers a pub/sub architecture, \textit{Media-over-QUIC Transport (MoQT)} (§\ref{sec:moq}), to carry DNS RRs and devise a strawman system and protocol proposal in §\ref{sec:design}.
Following, we provide a prototype implementation, and discuss our learnings and open challenges in §\ref{sec:evaluation}; we then conclude our paper in §\ref{sec:conclusion}.

\section{On the dynamics of today's DNS}
\label{sec:ttls}

To contextualize our motivation for a publish-subscribe variant of DNS, we first analyze the dynamics of today's DNS in terms of the usage of RRs in the Internet, their change rate over time, and the distribution of their TTLs.
We focus on the most commonly used record types on stub to recursive resolver requests, namely \textit{A} (IPv4 addresses) and \textit{AAAA} (IPv6 addresses).
In addition, we also analyze the 2024-standardized \textit{HTTPS} record type~\cite{rfc9460} which, amongst others, signals \textit{Application-Layer Protocol Negotiation (ALPN)} support within DNS.

\begin{figure}[t]
    \centering
    \begin{subfigure}{1\columnwidth}
        \includegraphics[width=\linewidth,clip,trim=35 7 50 40]{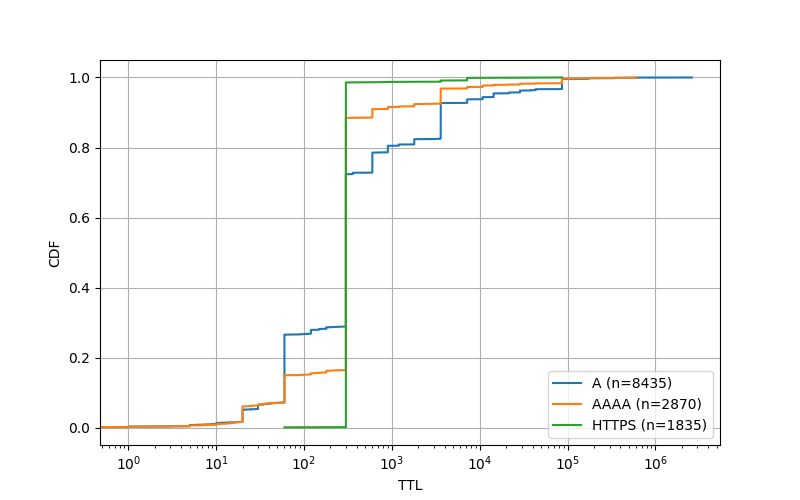}
        \vspace{-1.5em}
        \caption{TTL distribution}
        \label{fig:ttl}
    \end{subfigure}
    \begin{subfigure}{1\columnwidth}
        \includegraphics[width=\linewidth,clip,trim=35 7 50 0]{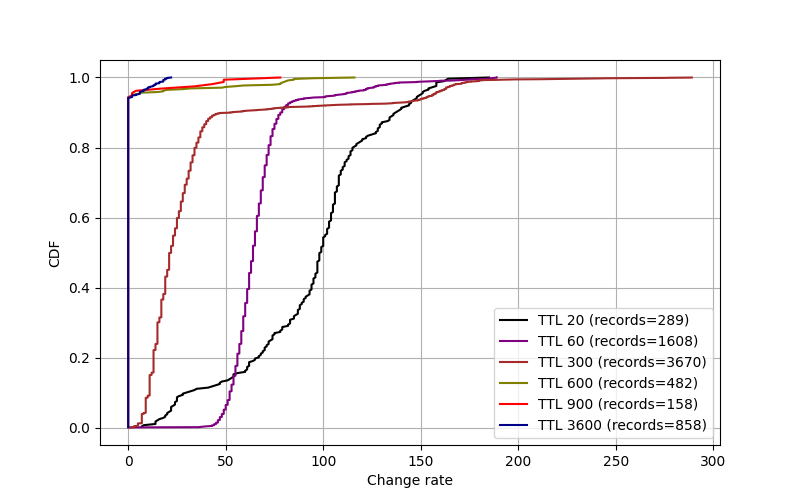}
        \vspace{-1.5em}
        \caption{\textit{A} record change rate over 300 observations}
        \label{fig:ttl_combined}
    \end{subfigure}
    \vspace{-2em}
    \caption{TTL distribution and \textit{A} record change rate}
    \vspace{-1em}
    \label{fig:ttl_analysis}
\end{figure}

Using the Tranco toplist from 2025-06-24, we recursively resolve the top 10k domains from a single vantage point in central europe, thereby resolving 8435 \textit{A} records, 2870 \textit{AAAA} records, as well as 1835 \textit{HTTPS} records as detailed in Fig.~\ref{fig:ttl}.
Surprisingly, the number of domains providing \textit{AAAA} records is only a fraction of those providing \textit{A} records, even more than 2 decades after the standardization of this record type.
In contrast, the uptake of the 2024-standardized \textit{HTTPS} record is astonishing, and highlights yet again the ever ongoing evolution of the DNS.

Analyzing the TTLs of the resolved records, we find that while each record type shows a distinct distribution, they naturally cluster in the TTLs [20, 60, 300, 600, 1200, 3600]\,s for \textit{A} and \textit{AAAA} records; notably, \textit{HTTPS} records are observed almost exclusively with a TTL of 300\,s, warranting a future investigation.
Based on these clustered TTLs, we next quantified the number of changes per record type over 300 subsequent observations of the respective TTL; e.g., for records with a TTL of 30, we performed recursive lookups every 30 seconds for a timeframe of 30\,s*300=150\,min.
We next compared the lexicographic ordered sample on positions $n$ to $n-1$, observing if the record changed within the TTL.
With the comparison in a lexicographic order, we counter the possible bias of DNS round-robin where the same entries per record type are responded in subsequent requests, but in a different order.

As detailed in Fig.~\ref{fig:ttl_combined} for \textit{A} records, we find that the lower the TTL the more changes are performed: while TTLs of 300\,s and below show a high change rate with at least 71 changes in the 90th percentile over 300 subsequent observations, TTLs of 600\,s and above show no changes at all up to the same percentile.
We make the same observations for \textit{AAAA} records; as for \textit{HTTPS} which are almost exclusively found with a TTL of 300\,s, we find that the number of changes is similar to \textit{A} records with a TTL of 300\,s.

As a key takeaway, we conclude that the dynamics of today's DNS are twofold: while the change rate of lower TTLs is rather high, the change rate of higher TTLs is rather low.
As such, we find that the possible benefits a publish-subscribe variant of DNS can provide is also twofold: first, it reduces the number of RR requests since updates are pushed to the subscribed resolvers, thereby limiting update traffic.
Second, the distribution of updates ensures that the subscribed resolvers are always up-to-date with the latest version of the record, even if the TTL has not yet expired; as such, a publish-subscribe variant of DNS can considerably reduce the time it takes for a resolver to receive the latest version of a record.

\section{Media-over-QUIC}
\label{sec:moq}

Media-over-QUIC is a framework to enable scalable media delivery which is currently being developed by the IETF.
The core of the framework is the publish-subscribe protocol Media over QUIC Transport
(MoQT)~\cite{ietf-moq-transport-12}, an application layer
protocol responsible for delivering media.

Although MoQT is developed with media use cases in mind, it is intentionally kept generic to enable non-media use cases and thus a suitable candidate for a publish-subscribe implementation of DNS that also provides a reliable and encrypted transport protocol by default. Other publish-subscribe protocols might be suitable as well and could be evaluated as alternatives to MoQT in the future.

Other components of the MoQ framework
include \textit{container formats} for encapsulating media segments for
transmission over MoQT, \textit{catalogs} for signaling availability of
collections of media tracks and their parameters, and \textit{streaming formats}
describing how to combine the individual MoQT components to build applications. 
To define a DNS over MoQT protocol, one needs to define a \textit{container format} to
encapsulate DNS messages in MoQT and define a \textit{streaming format} describing how applications such as nameservers and resolvers can use DNS over MoQT.

The MoQT protocol defines several control messages
for signaling and multiple messages for data delivery over QUIC. All control
messages in MoQT are exchanged using a single bidirectional QUIC stream, while
all object messages are sent over either unidirectional QUIC streams or
unreliable datagrams~\cite{rfc9221}. The control messages can be used to encode DNS requests, while DNS responses can be carried in MoQT's data messages.
MoQT uses a combination of namespaces and track names to uniquely identify
tracks. Namespaces are defined as a tuple of sequences of bytes, and a track name
is a single sequence of bytes. The maximum total length of the combination of
namespace and trackname is allowed to be 4096 bytes, which gives enough space to encode DNS requests.
The MoQT control messages include messages for session establishment and setup
of publications and subscriptions of tracks. In addition to live subscriptions,
MoQT also supports fetching already existing objects, e.g., objects from past
live streams. Data in MoQT is grouped into tracks, and tracks are further divided into groups of
objects. This model fits the media use cases well, e.g., video streams can be
split into groups of pictures, each containing multiple frames (objects), but since the objects are agnostic of the data they contain, they can also carry DNS response messages.

A goal of the MoQ framework is high scalability, supported by third-party relay providers. Relays are MoQT endpoints that do not publish or consume media but forward and route objects from publishers to subscribers. Relays can aggregate subscriptions of multiple subscribers to a single upstream subscription and cache objects without accessing the object payload. Since relays work on objects irrespective of the
content, they can also forward objects carrying DNS messages, which supports distributing the load of DNS message updates from authoritative servers to stub and recursive resolvers.


\section{Design}
\label{sec:design}

We will now detail our proposal for a mapping of DNS to MoQT, introducing MoQT 
as a protocol for communication between stub resolvers,
recursive resolvers, and authoritative nameservers.
DNS clients using MoQT connect to the resolver or nameserver
using QUIC and subscribe to updates for the record using the process detailed
in §\ref{subsec:lookup-operation}. Authoritative nameservers supporting MoQT
need to publish updates to subscribers using the update operation which is described in
§\ref{subsec:update-operation}. §\ref{subsec:message-formats} details
how we mapped DNS messages to MoQT. §\ref{subsec:teardown} describes
considerations for cleaning up subscriptions when they are no longer needed and
§\ref{subsec:fallback} explains a fallback mechanism for compatibility with the
traditional DNS.

\subsection{Lookup Operation}
\label{subsec:lookup-operation}

\begin{figure}[t]
    \begin{center}
    \includegraphics[width=1.0\columnwidth]{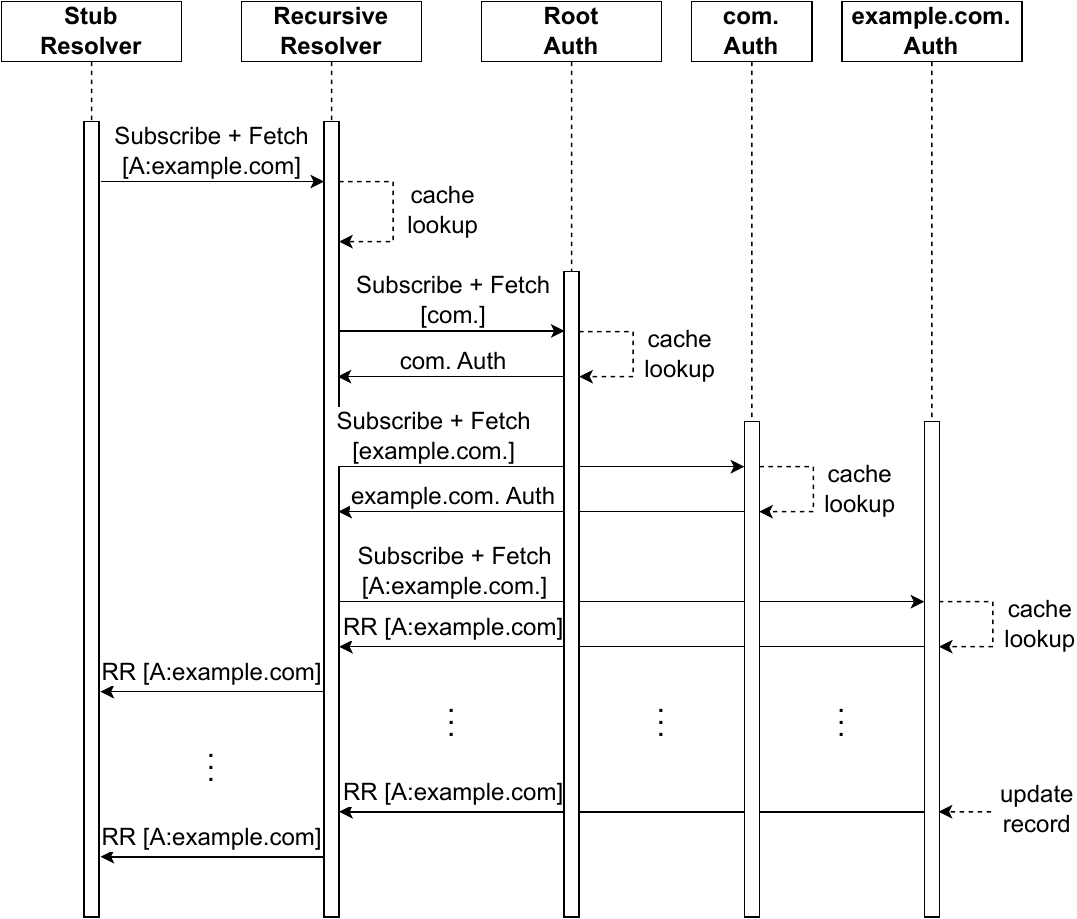}
    \end{center}
    \caption{Recursive DNS over MoQT Lookup Sequence}
    \label{fig:lookup-sequence}
\end{figure}

The task of a recursive resolver is to recursively look up DNS records,
starting with requesting the TLD server for the requested domain from one of the
root servers, then requesting the authoritative nameserver for the
record, and finally requesting the record from the identified
authoritative nameserver. DNS over MoQT does not change the
recursive nature of the process. Instead of using requests and responses, the
lookup over MoQT uses the MoQT procedures to fetch and subscribe to data as detailed
in Fig.~\ref{fig:lookup-sequence}.

To look up the latest version of a DNS record and subscribe to updates for the
record from an upstream server, a resolver opens a QUIC connection to
the server, indicating MoQT as the application layer protocol. After the
connection is established, a MoQT handshake establishes the MoQT session.
The resolver can now use the MoQT primitives
to fetch the latest version of the requested record and subscribe to further
updates of the record.

MoQT uses the control messages \texttt{FETCH} to fetch existing objects and
\texttt{SUBSCRIBE} to subscribe to objects generated in
the future. Combining both messages is called \text{joining fetch} and works by
first initiating a subscription, and then issuing a fetch-operation starting at
a relative offset before the start of the subscription. To request a record and
subscribe to updates for the record, resolvers start the
subscription for the requested record type and then fetch the version
immediately before the start of the subscription by using an offset of one.

The endpoint receiving the \texttt{SUBSCRIBE} and \texttt{FETCH} messages must
respond to each individually. To acknowledge and confirm subscriptions and
fetches, they use the \texttt{SUBSCRIBE\_OK} and \texttt{FETCH\_OK} control
message respectively. To signal an error in either of the operation, they use
the \texttt{SUBSCRIBE\_ERROR} and \texttt{FETCH\_ERROR} messages.

If the response is successful, the publisher opens a new QUIC stream and starts
sending DNS responses encapsulated in MoQT objects. Our DNS over MoQT prototype
uses QUIC streams and no datagrams to avoid losing messages due to the
unreliability of datagrams.

\subsection{Update Operation}
\label{subsec:update-operation}

Whenever a record is updated on an authoritative nameserver, it pushes the
updates to all clients that are connected via MoQT and have subscribed to
updates for the record. The update is pushed as a new MoQT object containing the
updated response for the request that opened the subscription.

In MoQT, objects are identified within a track by their group and object IDs. If
two objects within the same track have the same group and object IDs, their
content must be exactly the same.
Thus, if two subscribers subscribe to the same namespace and track name, they
both should receive the same objects. If one of them starts the subscription at
a later time, it may miss some of the earlier versions, but every object with
the same group and object ID should look the same for both subscribers.


To generate object IDs, authoritative servers in DNS over MoQT keep a version
number of the managed zone. The version number is a strictly monotonically
increasing sequence of integers. When a record in the zone changes, the version
number is increased, and an update is sent to all subscribers who are
subscribed to a track that includes the updated record in its answer message.
The server then generates a new answer message for each of the tracks and sends
it in an object with the group ID set to the new version number and the object
ID set to zero.


\subsection{Message Formats}
\label{subsec:message-formats}

\begin{figure}[t]
    \begin{center}
    \includegraphics[width=1.0\columnwidth]{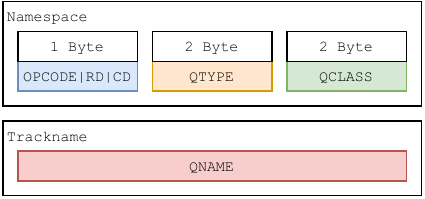}
    \end{center}
    \caption{DNS Query to MoQT namespace and trackname mapping}
    \label{fig:encapsulation-trackname}
\end{figure}

As detailed in §\ref{sec:moq}, MoQT uses namespaces and track names to
identify tracks. MoQT publishers can fan-out messages to multiple subscribers,
if all the subscribers are subscribed to the same track. To ensure that
different subscribers use the same combination of namespace and track name,
we map only the relevant fields of the DNS request message to
the MoQT namespace and track name as shown in
Fig.~\ref{fig:encapsulation-trackname}.

Specifically, we map five fields of the DNS request message to the first three elements of the namespace tuple. The first tuple element is a single byte including the four \texttt{OPCODE} bits, and one bit for the \texttt{RD} (Recursion Desired), and \texttt{CD} (Checking Disabled) fields each. The second tuple element is a two-byte field carrying the \texttt{QTYPE} field, and the third element is another two-byte field for the \texttt{QCLASS} field.
Additionally, we map the \texttt{QNAME} field of the DNS question section to the
MoQT track name field. Considering the limit of 4096 bytes for the combination of
namespace and track name, this mapping leaves 4091 bytes for the size of the
\texttt{QNAME} field.

\begin{figure}[t]
    \begin{center}
    \includegraphics[width=1.0\columnwidth]{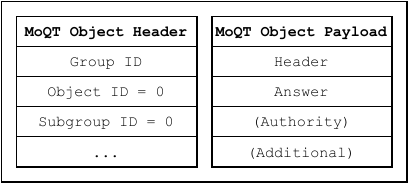}
    \end{center}
    \caption{DNS response encapsulation in MoQT objects}
    \label{fig:encapsulation-objects}
\end{figure}

Payload data in MoQT is sent in objects. Next to the payload, objects contain a
set of metadata fields including a \textit{group ID} and an \textit{object ID}.
We map response messages to MoQT by mapping the full DNS response message to the
payload field of the object, as shown in Fig.~\ref{fig:encapsulation-objects}.
Since there is no concept of grouped objects in DNS over MoQT, the object ID is
always set to zero, and the group ID is set to the version number introduced in
§\ref{subsec:update-operation}. Groups in DNS over MoQT always contain only one
object.

\subsection{Subscription Teardown}
\label{subsec:teardown}

Resolvers need to implement additional state management for MoQT subscriptions.
A clean-up routine for subscriptions that are no longer useful is required to
avoid wasting resources on unused subscriptions. Keeping subscriptions open for
a long time also has privacy implications since it leaves a trail of the
domain names that a client was previously interested in. Stub resolvers running on end-user devices also need to clean up subscriptions after suspension or shutdowns. Subscriptions can be re-established after the client device reconnects using the fetch mechanism described above. Stub resolvers can store the last known group ID of the corresponding subscriptions, and on reconnection, they can fetch any updates starting from the last known group ID by fetching and subscribing to objects following that ID.

The timescale at which resolvers can drop unused subscriptions depends on a trade-off between the acceptable overhead of managing the MoQT session and subscription state, and the risk of having to re-establish a new session and subscription if the record is requested again in the future. It is up to the resolver implementation to define the dynamics of the clean-up procedure, which could also be dynamically adapted based on the history of how frequently a domain had to be resolved in the past and how likely it is to be requested again in the future.


\subsection{Compatibility}
\label{subsec:fallback}

To allow incremental deployments of DNS over MoQT, recursive resolvers
need to provide their service via MoQT and any of the traditional DNS protocols.
Stub resolvers can then choose to keep using traditional DNS, or migrate
to MoQT~\cite{rfc9462}.
Compatibility also needs to be ensured when a stub resolver tries to
open a subscription with a recursive resolver for a domain that is
managed by an authoritative server that does not support MoQT.

If the recursive resolver has not yet established a MoQT session to the
authoritative server, and unless it has out-of-band information about the
supported protocols of the authoritative server, the resolver does not know if
the server supports MoQT. In that case, the resolver can use a happy
eyeballs-like approach by trying to establish a MoQT connection while
simultaneously sending a request over UDP.

If the authoritative server does not support MoQT, the recursive resolver
forwards the response to the traditional DNS to the stub resolver. Since the
recursive resolver does not receive any updates, it can only respond to the
\texttt{FETCH} request with the record received from the authoritative server.
To decline the subscription, the recursive resolver uses the
\texttt{SUBSCRIBE\_ERROR} message to decline the \texttt{SUBSCRIBE} request.

Alternatively, the recursive resolver can provide updates to the subscription by
periodically re-requesting the record from the authoritative server.
Periodically fetching new records puts more overhead on the recursive
resolver, but the interval can be reduced to the duration of the TTL, as that
is also the interval at which updates would at most be requested from the
authoritative server using traditional DNS over UDP.

\section{Discussion}
\label{sec:evaluation}

To facilitate a substantiated discussion of DNS over MoQT, we built a 
prototype implementation of the design described in §\ref{sec:design}.
The implementation is built on top of the
\texttt{mengelbart/moqtransport}~\cite{moqtransport}, \texttt{quic-go/quic-go}~\cite{quic-go}, and
\texttt{miekg/dns}~\cite{miekg-dns} libraries. It includes an
authoritative nameserver, a recursive resolver, and a
forwarder. The forwarder only forwards DNS requests to recursive
resolvers using MoQT. The recursive resolver uses the process
outlined in §\ref{subsec:lookup-operation} to resolve DNS records and subscribes to updates from authoritative servers.
The recursive resolver can respond to stub resolver requests using traditional
DNS protocols or MoQT. While we have not implemented a standalone MoQT stub resolver yet, 
the forwarder can provide DNS over MoQT functionality directly at the client when being operated on the same device, thereby also enabling backwards compatibility with traditional DNS stub resolvers.

Our prototype implementation shows the inherent benefits of a publish-subscribe model for the DNS in terms of update time where we find that the time it takes for a resolver to receive the latest version of a record can be considerably reduced depending on the actual TTL.
Moreover, the number of RR requests is reduced since updates are pushed to the subscribed resolvers, thereby effectively limiting update traffic.
The following sections will now detail our learnings thus far, highlighting open challenges and possible optimizations of the design choices.



\subsection{State Management Overhead}

DNS over UDP is stateless in the sense that it does not require endpoints to
keep connection state. In contrast, MoQT requires endpoints to permanently manage connection state. While DNS over TLS, HTTPS, or QUIC also need to manage connection state, DNS over MoQT adds the MoQT session and state for every open subscription.

Keeping long-running QUIC connections also requires endpoints to regularly test
the liveness of the connection. While connections can in theory be kept open in
idle state for long times (governed by the QUIC \texttt{max\_idle\_timeout}
transport parameter), endpoints should regularly test the liveness of the
connection. If a connection is silently closed, a subscriber risks missing
updates for a track and having to restart the QUIC connection, MoQT session, and
subscription when the next lookup is initiated.

\subsection{Query Latency}

The query latency of DNS mostly depends on the protocols used to transmit
requests and responses, and the state of the caches of intermediary resolvers.
Absent of any packet loss or timeouts, a recursive resolver can resolve a name from an authoritative server in a single round-tip. If it uses DNS over MoQT, and
there is no connection to the server, it takes at
least three round-trips: one round-trip for the QUIC connection, one for the MoQT session establishment, and one for the MoQT subscription. If the recursive
resolver needs to set up connections to multiple authoritative servers, possibly
including root and TLD servers, this process adds a considerable overhead to the
query latency observed by a stub resolver.

The overhead can be reduced with three optimizations. First, the recursive
resolver can reuse QUIC connections and the associated MoQT sessions for multiple
subscriptions. For some servers, such as root servers, the recursive resolver can
likely keep the session open permanently, which reduces the number of
round-trips to look up a record to a single round-trip, which is thereby on par with DNS over UDP. Second, if the resolver
had opened a connection to an authoritative nameserver in the past, it can use
QUIC's 0-RTT feature. 0-RTT allows sending application data in the first
round-trip, which reduces the number of round-trips necessary to establish a MoQT
session down to a single round-trip. The third optimization is not possible with
the version of the MoQT protocol as of the time of writing, but will likely be
possible in the future. The current version of MoQT requires a handshake to
establish the session before any further MoQT messages can be processed. The
handshake includes a version negotiation that needs to be finished before the
session can be used. Future versions will most likely move the version negotiation
to the transport protocol handshake, e.g., using Application Layer Protocol Negotiation
(ALPN)\footnote{\url{https://github.com/moq-wg/moq-transport/pull/499}}.

For stub resolvers, similar considerations apply. If the stub resolver has to
establish a QUIC connection, a MoQT session, and a subscription to resolve a name,
the latency will be much higher than using traditional DNS over UDP. However, if it can
reuse existing connections, the latency can be reduced to a single round-trip to
the recursive resolver. A bigger advantage can be achieved if the stub resolver
automatically receives updates for frequently used domains via MoQT. In this
case, the application does not have to make any lookup via the network at all.
Browsers, for example, could start loading a requested page immediately without
looking up a domain name first, which otherwise increases the page load time by a few milliseconds.

\subsection{Use Case Discussion}

We now come back to our three sample uses from §\ref{sec:intro}.  Dynamic DNS (DDNS) users, or their ISPs on their behalf, could directly publish RR updates to their associated authoritative name server; the MoQT infrastructure would then take care of update distribution. Since the propagation is limited to those subscribed to the respective domain name -- and servers hosted in home networks usually see limited interest
-- we expect the update traffic to be rather low.  Since the number of DDNS users is hard to estimate, let's do this by example: Let us assume 100M users worldwide with 1,000 other users each interested in their hosted services and involving 5 MoQ relays on average.  At two IP address updates per day and 300\,B update size, this would yield a globally distributed application layer update traffic of some 5.5\,Gbps, which is negligible at global scale.

CDN providers can also simply push their updates to their downstream users, i.e., recursive and stub revolvers.  Due to the typical popularity of CDN-hosted sites, one may expect that no excess traffic is generated towards the recursive resolvers (at least not during busy hours); rather to the contrary, the DNS requests flowing in the opposite direction are avoided.  Only stub resolvers may see extra traffic as users usually won't continuously request all the sites they are normally interested in: the average user may visit 100+ web sites per day (not necessarily all different ones) and close to 1,000 per week.\footnote{\url{https://www.digitalsilk.com/digital-trends/top-website-statistics/}}  Conservatively assuming that a stub resolver subscribes to 1,000 different domains and all domains are updated at the lowest observed clustered TTL of 10\,s with 300\,B per update, we obtain a downstream update traffic of 240\,kbps.  As discussed above, stub resolvers may tune their retention period for subscriptions according to the needs and limitations of access networks and/or user devices.

Finally, a deep space network could benefit from the same push mechanisms to update domain information on other planets, especially if the domains primarily used are well-managed to not provide excess update traffic.  Forwarding of records for domains observed to provide high update rates could be throttled since, e.g., load balancing for the closest and fastest responding CDN node won't make much of a difference for deep space access.
MoQT connections could supposedly be established to deep space nodes by deploying dedicated routers and applying suitable transport layer adaptations that can cope with long communication delays and disruptions as discussed for QUIC in \cite{many-tiptop-quic-profile-00}.


\section{Conclusion}
\label{sec:conclusion}

Our work shows that the DNS can benefit from a publish-subscribe variant. As such, DNS over MoQT ensures that resource records are always up to date on subscribed resolvers, countering the TTL-based request-response style lookups in the traditional DNS.
On the other hand, the new architecture also introduces new challenges, such as higher state management overhead on resolvers and nameservers, and increased query latencies for the first lookup of a record caused by higher connection and session establishment latencies. Future work is required to provide a more extensive analysis of a deployment of the new architecture.


\bibliographystyle{ACM-Reference-Format}
\bibliography{main}

\end{document}